\documentclass[epj]{svjour}
\usepackage{graphics}
\begin{document}

\title{Doping Effects on the Charge-Density-Wave Dynamics \\
in Blue Bronze}
\author{S. Yue\inst{1} \and C. A. Kuntscher\inst{1}\thanks{email: kuntscher@pi1.physik.uni- stuttgart.de} \and M. Dressel\inst{1}
\and S. van Smaalen\inst{2} \and F. Ritter\inst{3} \and W. Assmus\inst{3}}

\institute{1. Physikalisches Institut, Universit\"at Stuttgart,
Pfaffenwaldring 57, D-70550 Stuttgart, Germany \and Lehrstuhl
f\"ur Kristallographie, Universit\"at Bayreuth, 95440 Bayreuth,
Germany \and Physikalisches Institut, Universit\"at Frankfurt,
60054 Frankfurt, Germany}

\date{Received: \today}

\abstract{The temperature dependences of the dc resistivity and the nonlinear transport properties in pure,
Rb-doped, and W-doped blue bronze K$_{0.3}$MoO$_{3}$ single crystals are presented. In comparison with the
Rb doping, the W doping has larger effects on the electrical transport properties and the Peierls
transition. In particular, the maximum in the temperature dependence of the threshold field for nonlinear 
transport, observed in pure and Rb-doped samples around 100 K, is absent in W-doped K$_{0.3}$MoO$_{3}$.
These results are discussed with respect to the proposed incommensurate-commensurate transition of the 
charge-density-wave and its interaction with impurities and normal carriers.
\PACS{
      {71.45.Lr}{Charge-density-wave systems} \and
      {71.30.+h}{Metal-insulator transition and other electronic transitions}
      }
} 
\maketitle

\section{Introduction}
Charge-density-waves (CDW) in low-dimensional solids attract
interest since R. Peierls introduced this broken-symmetry ground
state half a century ago \cite{peierls55}. Their dynamics has been
subject to numerous investigations as soon as the first
experimental realization became possible in the 1970s. The present
state of knowledge is compiled in several reviews
\cite{monceau85,gruner85,gruner88,schlenker89,gorkov89,gruner94}.
Below the Peierls transitions, an energy gap opens at the Fermi
energy leading to an insulating ground state. In the CDW phase the
electronic transport has contributions from the normal charge
carriers excited across the gap (like a semiconductor) and from
the collective response of the CDW condensate. Although there
exist a general idea about the main phenomena, a large number of
observations are still puzzling and the details are not understood
in depth. One of the key issues of the CDW dynamics is the
interaction of the condensate with the underlying lattice and in
particular with lattice imperfections.

Molybdenum bronze $A_{0.3}$MoO$_{3}$, where $A$ = K, Rb or Tl, is
among the most-extensively studied quasi-one-dimensional
compounds, which develops a CDW along the $b$ axis below approximately 180~K
\cite{schlenker89}. In contrast to ideal systems the CDW
condensate cannot move freely, but is pinned to the lattice by
impurities and defects. This is easily seen, for instance, if an
electric field is applied along the chain direction exceeding a
critical value, the so-called first threshold field $E_{T}$; as a
consequence the CDW is depinned, it begins to slide through the
crystal and contributes additionally to the conductivity,
resulting in a nonlinear current-voltage ($I-V$) response \cite{dumas83}. 
Below approximately $T=40$~K for K$_{0.3}$MoO$_{3}$ a more complicated
$I-V$ response is observed, including a second threshold field $E_{T}^*$
above which a sharp increase of the current by many orders of magnitude
occurs \cite{mihaly87,littlewood88,martin88}.
The temperature dependence of the first threshold field
$E_{T}$ is commonly explained by a sort of two-fluid model
containing the electrons condensed in the CDW and the remaining
normal electrons \cite{sneddon84}; the approach describes the
Coulomb interaction between the CDW and uncondensed charge
carriers thermally excited above the Peierls gap. Within this
model, with rising temperature the uncondensed carrier density
increases, leading to an increasing damping of the CDW by dissipative
normal currents and thus to an increase of $E_{T}$.
However, when the temperature rises above $T=100$~K a
decrease of $E_{T}$ is observed, resulting in a maximum of threshold field
$E_T(T)$ around this temperature \cite{dumas83}. Similar
observations were reported for the CDW compound TaS$_3$
\cite{fleming86}. The change in the temperature dependence of
$E_T(T)$ might be attributed to an incommensurate-commensurate (I-C)
transition of the CDW which happens around $T=100$~K for blue
bronze. However, small deviations from complete commensurability
were found down to 4~K \cite{fleming85,sato83,pouget85,janossy88}
which somehow disturbs this picture. 
Interestingly, in comparison with K$_{0.3}$MoO$_{3}$ and TaS$_3$, 
for NbSe$_3$ with two CDWs, which are always incommensurate with the 
underlying lattice, a very different temperature dependence of $E_{T}(T)$ 
was found \cite{fleming80}: The threshold field diverges at the 
Peierls transition temperature $T_P$ and exhibits a minimum slightly
below $T_P$. This behavior could be well described by the 
Fukuyama-Lee-Rice model for impurity pinning of the CDW \cite{lee79,fukuyama78} 
taking into account thermal fluctuations of the CDW phase \cite{maki89}.
For K$_{0.3}$MoO$_3$ different groups observed the maximum in
$E_{T}(T)$ near $T=100$~K, but its origin remains a puzzle.

The peculiarities of the CDW motion may be clarified by studying
the doping dependence of $E_{T}(T)$. In blue bronze two types of
substitution are possible, namely replacing K by isoelectronic Rb
and the substitution of Mo by isoelectronic W. The two doping 
channels have a very different effect on the transport properties
\cite{schneemeyer84,schneemeyer85}. In blue bronze the oxygen
octahedrons with the transition metal in the center form
conducting chains along which the CDW develops. Obviously W
substitution strongly disturbs the one-dimensional conduction and
the development of the CDW, hence already a very small W
concentration causes a large shift of the Peierls transition to
lower temperature and a considerable broadening of the phase
transition. For this
so-called {\it strong} pinning the phase of the CDW is adjusted at
each impurity site \cite{lee79}. In contrast, the Rb dopants
occupying the K sublattice only cause some disorder in the crystal
potential and thus only play a minor role in the pinning mechanism; 
it is called weak pinning \cite{lee79}. 
Measurements of $E_{T}(T)$ of K$_{0.3}$MoO$_3$ with 50\% Rb doping 
for temperatures below 80 K show a similar behavior as in the pure 
compound \cite{cava85}. However, the temperature dependence of the 
first threshold field for other Rb doping levels, and in particular 
for W-doped blue bronze compounds has not been measured up to now.
Here we present a comprehensive study of $E_{T}(T)$ for a series of 
Rb- as well as of W-doped blue bronze compounds and discuss the 
origin of the maximum threshold observed around $T=100$~K.

\section{Experiments and Results}

Pure, Rb-doped, and W-doped blue bronze K$_{0.3}$MoO$_{3}$ single
crystals (see Table \ref{tab:Tp}) were prepared by the temperature gradient 
flux method \cite{ramanujachary84} and electrolytic reduction of the molten salts 
of $A_2$CO$_3$-MoO$_3$ for pure blue bronze $A_{0.3}$MoO$_3$ ($A$=K and Rb), 
K$_2$CO$_3$-Rb$_2$CO$_3$-MoO$_3$ for Rb-doped K$_{0.3}$MoO$_3$, and K$_2$CO$_3$-MoO$_3$-WO$_3$ 
for W-doped K$_{0.3}$MoO$_3$. Different concentrations of the dopants in the 
melt determine the doping levels in the obtained crystals. For Rb-doped  
samples, the doping ratio is almost the same as that in the melt, but for 
the W-doped compounds, the doping ratio is nearly 2 to 3 times bigger 
than that in the melt \cite{schneemeyer85,mingliang97}.
The obtained samples have a typical size of 4$\times$4$\times$0.5
mm$^{3}$. In our dc electrical transport measurements a four-probe
configuration was used. Thin gold wires were anchored by silver
paint. In the contact regions, indium films were covered by gold
films for stable contacting at low temperature. In this way, a
typical contact resistance of around $1~\Omega$ was achieved. For
samples with the same doping level but grown by different
preparation methods we obtained similar results.

\begin{table}[b]
\caption{List of the studied blue bronze samples with the Peierls
transition temperature $T_p$ and the width of the transition $\Delta$$T_P$
[full width at half maximum of the derivative $D(T)$].}
\label{tab:Tp}
\begin{tabular}{ccc}
\hline\noalign{\smallskip}
Sample & $T_p$ (K) & $\Delta$$T_P$ (K) \\
\noalign{\smallskip}\hline\noalign{\smallskip}
Rb$_{0.3}$MoO$_{3}$ & 180.5 & 7.3 \\
{(K$_{0.5}$Rb$_{0.5}$)}$_{0.3}$MoO$_{3}$ & 170.7 & 36.3 \\
{(K$_{0.833}$Rb$_{0.167}$)}$_{0.3}$MoO$_{3}$ & 175.1 & 18.5 \\
K$_{0.3}$MoO$_{3}$ & 179.2 & 11.2 \\
K$_{0.3}$Mo$_{0.995}$W$_{0.005}$O$_{3}$ & 160.7 & $>$ 50 \\
K$_{0.3}$Mo$_{0.99}$W$_{0.01}$O$_{3}$ & $\approx$140 & --- \\
K$_{0.3}$Mo$_{0.98}$W$_{0.02}$O$_{3}$ & $\approx$140 & ---
\\\noalign{\smallskip}\hline
\end{tabular}
\end{table}

\begin{figure}
\resizebox{0.99\columnwidth}{!}{\includegraphics{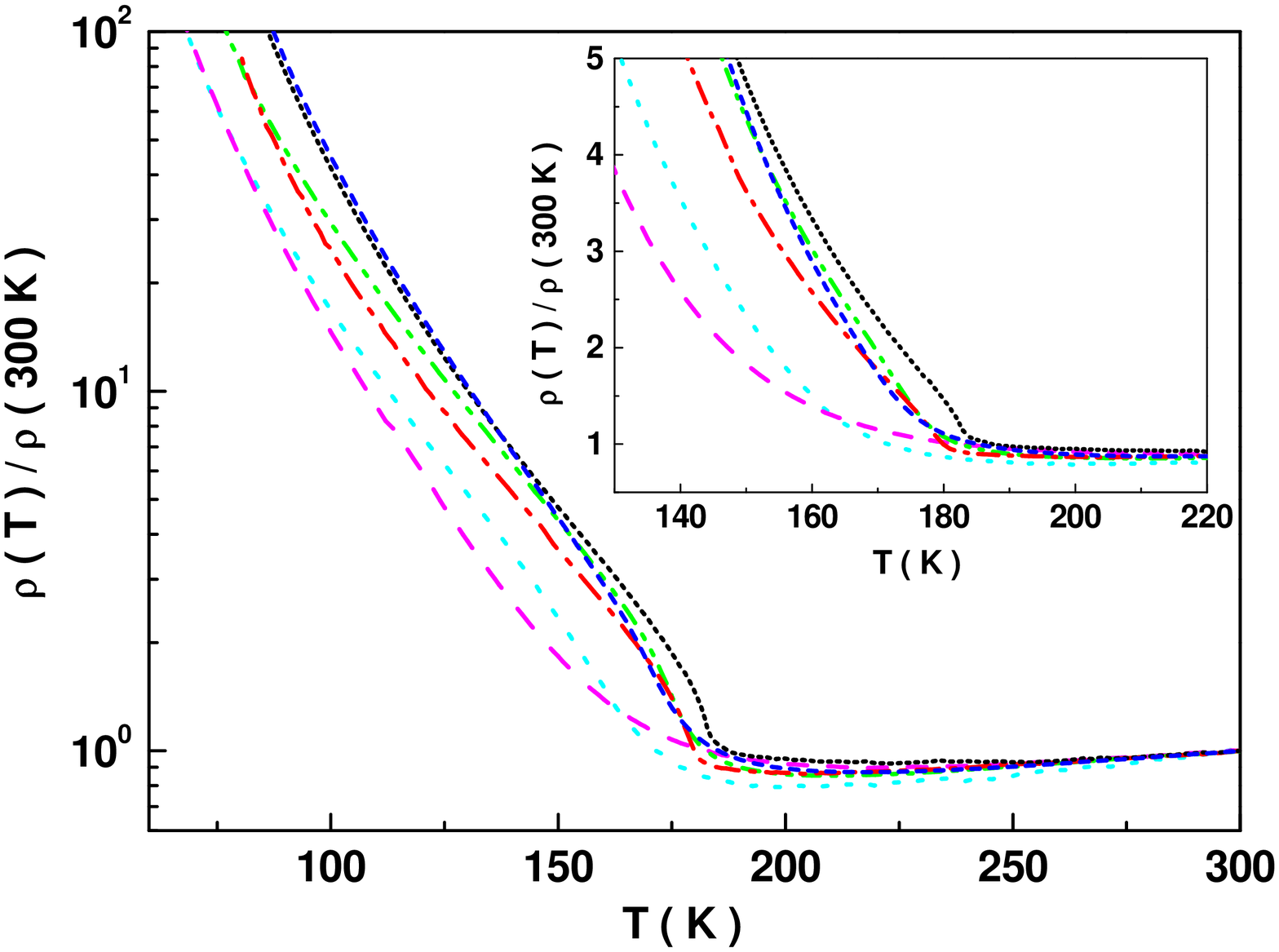}}
\caption{Dc resistivity along the \emph{b} axis, normalized to the
room temperature value, for 
Rb$_{0.3}$MoO$_{3}$ (black short dot),
{(K$_{0.5}$Rb$_{0.5}$)}$_{0.3}$MoO$_{3}$ (blue short dash),
{(K$_{0.833}$Rb$_{0.167}$)}$_{0.3}$MoO$_{3}$ (green dash dot dot),
K$_{0.3}$MoO$_{3}$ (red dash dot),
K$_{0.3}$Mo$_{0.995}$W$_{0.005}$O$_{3}$ (cyan dot) and
K$_{0.3}$Mo$_{0.99}$W$_{0.01}$O$_{3}$ (purple dash). 
The inset shows the enlargement of the results near the Peierls transition
temperature $T_p$.}
 \label{fig:RT}
\end{figure}

Fig.\ \ref{fig:RT} shows the temperature dependence of the dc
resistivity along the chain direction $b$ normalized to the room
temperature value for a series of pure and doped blue bronze
crystals. The pure samples K$_{0.3}$MoO$_{3}$ and
Rb$_{0.3}$MoO$_{3}$ show a sharp metal-semiconductor transition at
the Peierls transition temperature $T_p$. The mixed compounds with
16.7$\%$ and 50$\%$ substitution of K by Rb still exhibit a clear
transition $T_p$ which is slightly shifted to lower temperatures.
In contrast, for samples with small W doping (0.5$\%$, 1$\%$, and 2$\%$, 
whose results are very similar to those of 1$\%$ doping and therefore not
shown), the phase transition is significantly broadened. These
results are consistent with previous reports \cite{schneemeyer84}.

\begin{figure}
\resizebox{0.99\columnwidth}{!}{%
  \includegraphics{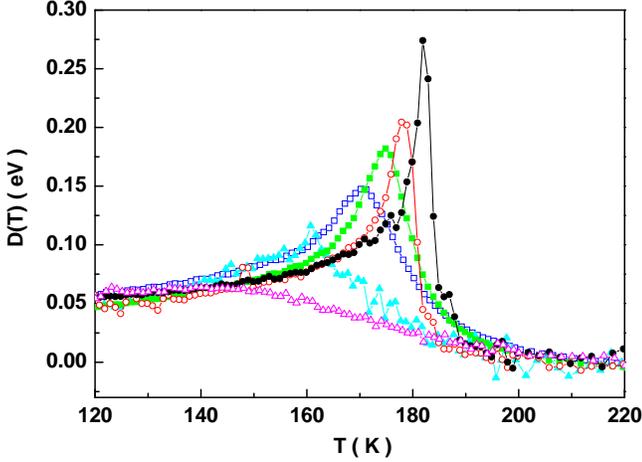}
} \caption{The derivative $D =
-\partial{\ln\sigma}/\partial{T^{-1}}$ versus temperature for
Rb$_{0.3}$MoO$_{3}$ (black full circles), K$_{0.3}$MoO$_{3}$ (open red circles), 
{(K$_{0.833}$Rb$_{0.167}$)}$_{0.3}$MoO$_{3}$ (full green
squares), {(K$_{0.5}$Rb$_{0.5}$)}$_{0.3}$MoO$_{3}$ (open blue squares),
K$_{0.3}$Mo$_{0.995}$W$_{0.005}$O$_{3}$ (full cyan triangles), and
K$_{0.3}$Mo$_{0.99}$W$_{0.01}$O$_{3}$ (open purple triangles).}
\label{fig:DT}
\end{figure}

To determine the Peierls transition temperature more precisely, we
plot the derivative $D(T) = -\partial{\ln\sigma}/\partial{T^{-1}}$
versus temperature in Fig.~\ref{fig:DT}. Except for the samples with 
1\% and 2\% (not shown) W doping, all curves clearly show a
peak around $T_P$; the width $\Delta$$T_P$ is a measure of the sharpness of the
transition \cite{carneiro85}. Both the Rb and the W doping lead to
a shift of the transition to lower temperature and a
broadening of the transition (see Table \ref{tab:Tp}); however,
these effects are considerably bigger in the case of
W substitution.

\begin{figure}
\resizebox{0.99\columnwidth}{!}{%
  \includegraphics{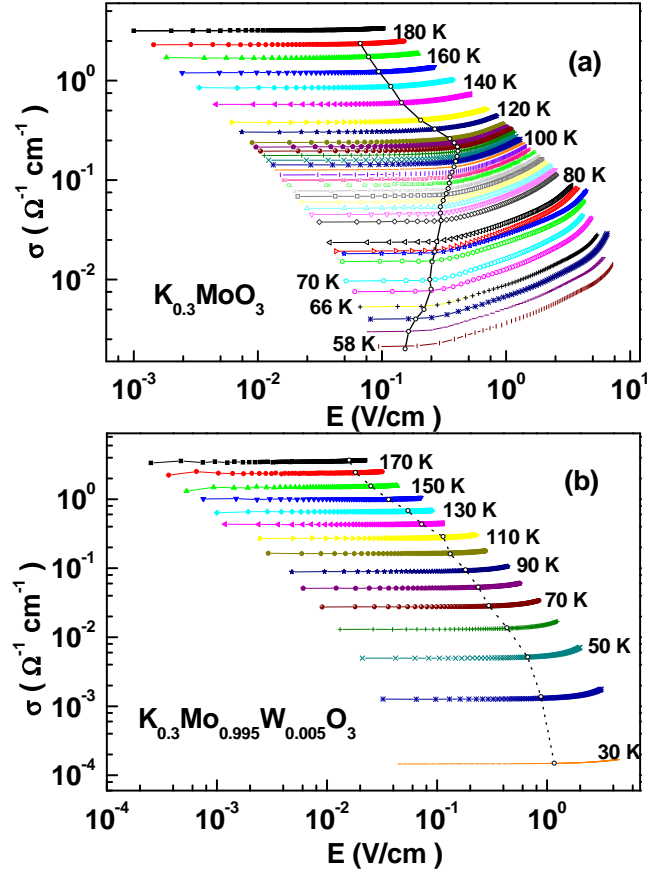}}
\caption{Dc conductivity along the \emph{b} axis as a function
of applied electric field for several temperatures below the
Peierls transition temperature $T_p$ for (a) K$_{0.3}$MoO$_{3}$
and (b) K$_{0.3}$Mo$_{0.995}$W$_{0.005}$O$_{3}$. The dashed lines
are guides to the eye for the variation of the threshold field
with temperature.}
 \label{fig:ET}
\end{figure}

\begin{figure}
\resizebox{0.95\columnwidth}{!}{%
  \includegraphics{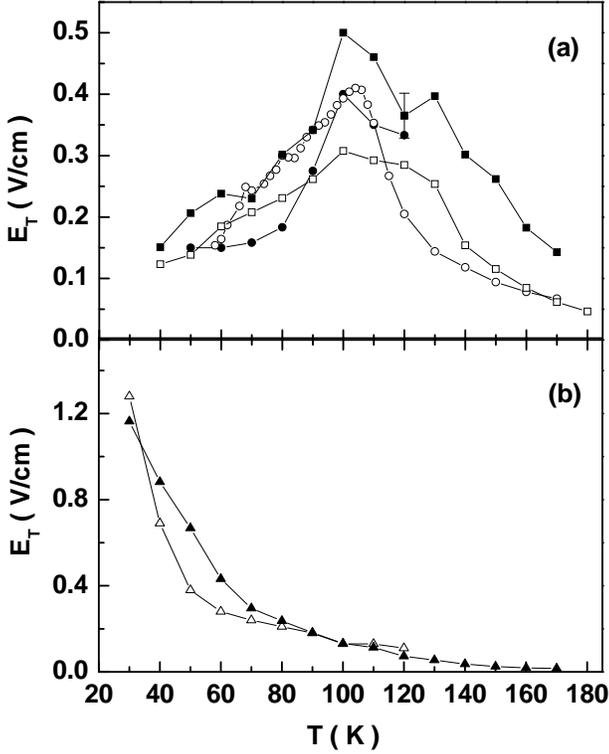}
} \caption{Temperature dependence of the threshold field $E_T$
along the $b$ axis for (a) Rb$_{0.3}$MoO$_{3}$ (full circles),
K$_{0.3}$MoO$_{3}$ (open circles),
{(K$_{0.833}$Rb$_{0.167}$)}$_{0.3}$MoO$_{3}$ (full squares), and
{(K$_{0.5}$Rb$_{0.5}$)}$_{0.3}$MoO$_{3}$ (open squares), and for
(b) K$_{0.3}$Mo$_{0.995}$W$_{0.005}$O$_{3}$ (full triangles) and
K$_{0.3}$Mo$_{0.99}$W$_{0.01}$O$_{3}$ (open triangles).}
\label{fig:ETT}
\end{figure}

We also studied the dc conductivity of pure and doped blue bronze
single crystals along the chain direction $b$ as a function of the
applied electrical field. Two examples are displayed in
Fig.~\ref{fig:ET}. For temperatures below $T_p$ the conductivity
increases rapidly when the applied electric field exceeds the
threshold field $E_T$ due to the depinning of the CDW. The values
of $E_T$  were determined graphically from the $I-V$
characteristics. Fig.~\ref{fig:ETT} summarizes the temperature
dependences of the threshold field for blue bronze samples with
different substitutions as indicated.

\section{Discussion}

The main results of our experiments can be summarized as follows:
(i)~Starting from the Peierls transition at $T_P=180$~K, for pure
K$_{0.3}$MoO$_{3}$ and Rb$_{0.3}$MoO$_{3}$ compounds $E_T$ first
increases upon cooling and then starts to decrease, resulting in a
maximum in the behavior of $E_{T}(T)$ around $T=100$~K; this
agrees with previous data \cite{dumas83}. 
(ii)~The substitution of K by Rb and of Mo by W does not influence the 
absolute value of $E_T$ significantly; in fact, for temperatures above 
$\approx$50 K the $E_T$ values for the W-doped samples 
are even slightly smaller than those for the pure samples, as
found previously \cite{cava85}. It is known that the absolute 
value of $E_T$ depends on the properties of the sample, 
such as impurity concentration, cross section, and current 
distribution.
(iii)~For all the Rb-doped samples the threshold field
$E_T(T)$ goes through a maximum value around $T=100$~K when the
temperature is lowered, like for the pure compound. However, even
for very small W substitution on the Mo sites, the temperature
dependence of the threshold field changed completely: $E_T(T)$
increases monotonously with decreasing temperature, and no maximum
is observed within our experimental temperature range.

The non-monotonous temperature behavior of $E_T(T)$
observed for the CDW phase in blue bronze is puzzling. The most
common explanation is the I-C transition
of the CDW around $T=100$~K. However, several observations 
contradict this hypothesis: If the CDW becomes commensurate with the
underlying lattice for temperatures below $T=100$~K, it should be strongly
pinned and thus $E_T(T)$ is expected to increase
below $T=100$~K, in contrast to the experimental findings.
Furthermore, substituting Mo by W adds pinning centers, but does
not change the CDW nesting vector and its temperature dependence \cite{fleming85};
nevertheless for W-doped compounds the maximum in $E_T(T)$ is absent.

Certainly, there is no abrupt transition from an incommensurate to
a commensurate CDW, but a continuous adjustment of the CDW to the
underlying lattice; maybe this process saturates around $T=100$~K
\cite{fleming85}. It is known that down to $T=4$~K small
deviations from perfect commensuration remain
\cite{sato83,pouget85,janossy88}. Results from electron spin
resonance experiments indicate the spatial coexistence of
commensurate and incommensurate regions; when lowering the
temperature from $T=53$~K to 25~K the commensurate fraction increases
from 5$\%$ to a value above 20$\%$ \cite{janossy88}. It was
proposed \cite{mcmillan76} that even the incommensurate CDW
consists of regions which are essentially commensurate with the
lattice separated by discommensurations; these latter are narrow
regions where the phase or displacement changes rapidly. At low
temperature the CDW will then not move as a rigid object but
propagate by depinned discommensurations, like the motion of 
grain boundaries and dislocations in the case of plastic deformation.

There are different effects relevant for the CDW transport, which
might also influence the temperature dependence of the threshold
field: (i)~The CDW can interact with the underlying lattice in
very different ways. Since for a commensurate CDW the interaction
with the lattice is strong, mainly incommensurate domains of
the CDW contribute to the collective transport. (ii)~The CDW is
not rigid, but subject to internal deformations, leading to a non-zero
threshold field for nonlinear transport. These dynamic deformations of 
the CDW and dissipative normal currents need 
to be taken into account in order to explain $E_T(T)$. 
(iii)~The interaction of the CDW with impurities or disorder either leads to
strong or weak pinning of the CDW \cite{lee79}. For strong pinning
the CDW distorts its phase and amplitude locally at each impurity
site; the CDW coherence length is therefore only on the order of
the impurity spacing. In the case of weak pinning, the CDW adjusts
its overall phase distribution within its coherence length, which
is much larger than the impurity spacing. While the substitution
of K by Rb pins the CDW only weakly, W doping causes strong
pinning of the CDW \cite{schneemeyer84}. 

The last two mechanisms will be strongly influenced by uncondensed electrons. 
According to Sneddon's model \cite{sneddon84} the presence of uncondensed
charge carriers will cause a damping of the CDW mode, since dissipative
normal currents are induced by the dynamic deformations of the CDW. This is
in agreement with previous investigations of the nonlinear characteristics
of pure blue bronze, where a strong influence of the single-particle 
conductivity on the collective CDW response had been found \cite{mihaly88}.
Within this picture $E_{T}(T)$ increases with rising temperature, as the fraction 
of uncondensed carriers increases. 
Also the strength of the interaction of the CDW with lattice imperfections,
like impurities or disorder, is influenced by the presence of normal charge
carriers. This is in particular important in the case of strong pinning, since
uncondensed carriers smoothen the CDW phase deformation induced by the 
strong pinning centers.

\begin{figure}
\resizebox{0.95\columnwidth}{!}{%
  \includegraphics{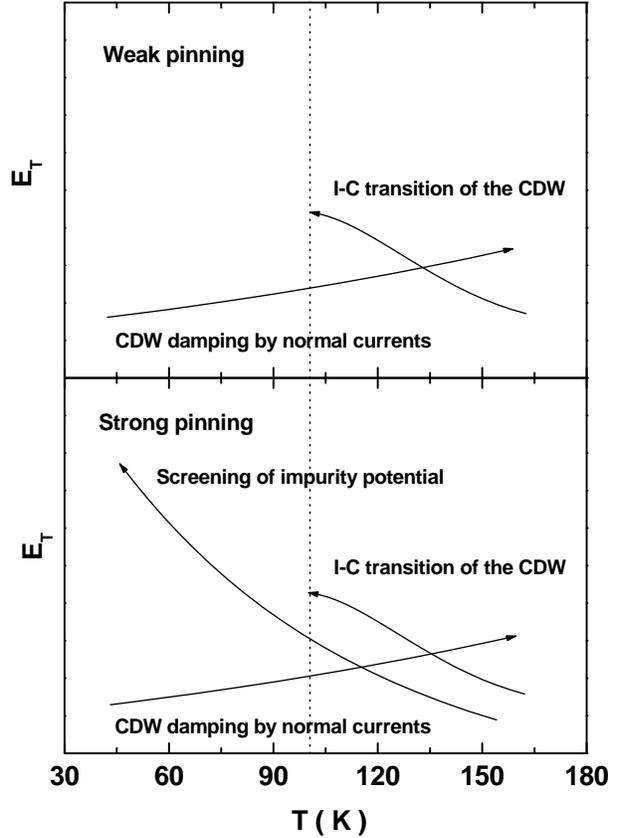}
} \caption{Scheme of the various mechanisms relevant for the
temperature dependence of the threshold field, $E_T(T)$, for weak
pinning (top) and strong pinning (bottom).} \label{fig:scheme}
\end{figure}

The temperature dependence of the threshold field $E_T$ is
determined by the above mentioned mechanisms, as illustrated in Fig.\
\ref{fig:scheme}. In the course of the I-C transition the CDW becomes
less discommensurate with decreasing temperature (with some sort of
saturation around 100~K), which causes $E_T(T)$ to increase. According
to Sneddon's two-fluid approach, the CDW damping by normal currents becomes
less important as the temperature is reduced, leading to a decrease of
$E_T(T)$. On the other hand, the screening of the impurity potential
by uncondensed carriers will also depend on their density and thus
on temperature, causing an increase of $E_T(T)$ as the temperature is 
lowered. This effect is expected to be only relevant for strong pinning,
i.e., for W doping in the case of blue bronze, but not for weak pinning,
i.e., Rb doping. This could explain why the maximum of $E_T(T)$ around 
$T=100$~K is observed only for the Rb-doped compounds, but not for W-doped 
samples. For the latter compounds a smooth increase of $E_T(T)$ with 
decreasing temperature is found.

\section{Conclusion}

In order to investigate the influence of strong and weak pinning on the
characteristics of the Peierls transition and on the first threshold field of
non-linear conductivity, we studied the electrical transport properties
of a series of pure, Rb-doped, and W-doped blue bronze
K$_{0.3}$MoO$_{3}$ single crystals. For pure K$_{0.3}$MoO$_{3}$
crystals and for substitution of K by Rb, which acts as weak pinning
center, the threshold field $E_T(T)$ exhibits a maximum around
$T=100$~K which is correlated with the
in\-com\-men\-su\-rate-to-com\-men\-su\-rate transition of the CDW. A
more advanced explanation requires to take into account the influence
of the uncondensed charge carriers on the CDW transport.

\section{Acknowledgements}
The authors would like to thank Kazumi Maki, Attila Virosztek, and Silvia Tomic 
for helpful discussion and G. Untereiner for technical support. Financial support
by the Deutsche Forschungsgemeinschaft (SPP 1073) is acknowledged.

\end{document}